\documentclass[paper,12pt]{JHEP}

\usepackage[centertags]{amsmath}
\usepackage{amsfonts} \usepackage{amssymb} \usepackage{amsthm}
\allowdisplaybreaks[1]
\usepackage{graphicx}
\usepackage{mathrsfs}
\usepackage[utf8x]{inputenc}
\usepackage{url}

\usepackage[]{amssymb}
\usepackage{bbold}

\newcommand{\ii}{\mathrm{i}}
\newcommand{\ee}{\mathrm{e}}

\newcommand{\one}{{\rm 1\kern -.9mm l}}

\newcommand{\ft}[2]{{\textstyle\frac{#1}{#2}}}

\newdimen\tableauside\tableauside=1.0ex
\newdimen\tableaurule\tableaurule=0.4pt
\newdimen\tableaustep
\def\phantomhrule#1{\hbox{\vbox to0pt{\hrule height\tableaurule
width#1\vss}}}
\def\phantomvrule#1{\vbox{\hbox to0pt{\vrule width\tableaurule
height#1\hss}}}
\def\sqr{\vbox{%
  \phantomhrule\tableaustep
\hbox{\phantomvrule\tableaustep\kern\tableaustep\phantomvrule\tableaustep}%
  \hbox{\vbox{\phantomhrule\tableauside}\kern-\tableaurule}}}
\def\squares#1{\hbox{\count0=#1\noindent\loop\sqr
  \advance\count0 by-1 \ifnum\count0>0\repeat}}
\def\tableau#1{\vcenter{\offinterlineskip
  \tableaustep=\tableauside\advance\tableaustep by-\tableaurule
  \kern\normallineskip\hbox
    {\kern\normallineskip\vbox
      {\gettableau#1 0 }%
     \kern\normallineskip\kern\tableaurule}%
  \kern\normallineskip\kern\tableaurule}}
\def\gettableau#1 {\ifnum#1=0\let\next=\null\else
  \squares{#1}\let\next=\gettableau\fi\next}
\def\XXint#1#2#3{{\setbox0=\hbox{$#1{#2#3}{\int}$}
     \vcenter{\hbox{$#2#3$}}\kern-.5\wd0}}

\def\be{\begin{equation}}
\def\ee{\end{equation}}
\def\bea{\begin{eqnarray}}
\def\eea{\end{eqnarray}}
\newcommand{\nn}{\nonumber}
\def\ii{{\rm i}}

\title{Black hole microstates from branes at angle 
}
\author{Lorenzo Pieri$^{1,2}$
\\
\vskip 0.2cm
$^1$ Universit\`a di Roma Tor Vergata and I.N.F.N, Dipartimento di Fisica, 
\\ Via della Ricerca Scientifica, I-00133 Rome, Italy\\
\vskip 0.2cm
$^2$ Centre for Research in String Theory,  School of Physics and Astronomy, \\
Queen Mary University of London, \\Mile End Road,  London, E1 4NS, United Kingdom \\
\vskip 0.2cm
\email{lorenzo.pieri@roma2.infn.it}
}
\abstract{We derive the leading $g_s$ perturbation of the SUGRA fields generated  by a supersymmetric configuration of respectively 1, 2 or 4 D3-branes intersecting at an arbitrary angle via the computation of the string theory disk scattering amplitude of one massless NSNS  field interacting with open strings stretched between the branes. The configuration with four branes is expected to be relevant for black hole microstate counting in four dimensions.}


\preprint{ROM2F/2016/07}

\begin{document}
\begin{Huge}

\end{Huge}
\pagebreak

\tableofcontents

\pagebreak

\section*{Introduction and Conclusion}

In a recent paper  was explicitly uncovered an open-closed dictionary between microstates in string theory arising from open string stretched between D-branes and microstate geometries in supergravity  for a system of 4 perpendicularly intersecting D3 branes \cite{Bianchi:2016bgx}. This configuration of branes  is particularly interesting, since it can be seen as the weak $g_s$ coupling picture of a four dimensional black holes with non vanishing area in the strong coupling (SUGRA)  picture and computation of its degeneracy should led to the microscopic calculation of the entropy \cite{Strominger:1996sh,Balasubramanian:1996rx}. 

We stress that for microstate geometries one means SUGRA solutions with the same asymptotic charges as the corresponding black hole, but differing from it near the location of the black hole horizon. The basic idea is that the singular black hole arises as a coarse-grained description of the underlying regular microstates when we trace out the region inside the would-be horizon, that is where the microstates differ from each other. The fact that these microstates  can be found directly in SUGRA is the statement of  the fuzzball proposal in a nutshell \cite{Mathur:2005zp}. More precisely, the strongest form of the conjecture says that  every string microstates is associated bijectively to regular and horizon-less microstates geometries. The beauty of this proposal lies in the possibility of solving the information loss paradox with tools already available, namely using mainly supergravity, branes and perturbative string theory. Of course a classical SUGRA solution  must be thought as a very special state (coherent state) of the real fully quantum picture, therefore in order to completely describe the features of a quantum black hole, a better knowledge of non perturbative string theory will be required. Nevertheless, we could hope to catch the relevant  physics in this approximation.

Much work has been done on 2 and 3 charge black holes, see for instance \cite{Mathur:2008nj}-\nocite{Bena:2007kg,Skenderis:2008qn, Giusto:2012yz, Bena:2011uw, Giusto:2012jx, Giusto:2013bda}\cite{Bena:2016ypk}, but very little is known for the four charge case, corresponding to black holes in four dimensions (apparently the dimension in which we live). In this paper we generalize the work done in \cite{Bianchi:2016bgx} by tilting the branes with some arbitrary angle, but still keeping the system supersymmetric. These configurations are relevant  if one wants to study orbifold compactifications rather than the simple toroidal ones, leading to more realistic black hole models. 

The strategy of the paper consists in deriving the leading $g_s$ perturbation of the SUGRA fields generated by the configuration of tilted branes via the computation of the string theory disk\footnote{The disk is the worldsheet surface associated to the leading $g_s$ contribution of a mixed open-closed amplitudes.} scattering amplitude of one massless NSNS  field interacting with open strings binding the different branes. We begin the stringy computations by evaluating the amplitude for the emission of one closed NSNS field in the presence of one single stack of branes (section 1). 
To compute the amplitude with two different stacks (section 3) bound by open strings we first construct the open string vertex operators associated to every relevant pair of branes taken into account (section 2), paying attention to be consistent with the supersymmetry preserved by the system. Finally we perform the calculation for the 4 brane system (section 4), that it's the novel result of this letter.

 Explicitly,  once the amplitude $ {\cal A}$ has been computed, to extract the leading perturbation is sufficient to use the effective field theory  formula

 \be
 \delta\tilde \Phi(k)
  =  \left( - {\ii  \over k^2} \right) {\delta {\cal A}(k) \over \delta \Phi} 
  \label{bulkb}
  \ee
  
In \cite{Bianchi:2016bgx}  explicit agreement was shown between the stringy calculations and a known extremal SUGRA solution, in line with the idea that these classical solutions can account for the microstates of the corresponding black hole. This was achieved by expanding the SUGRA solutions at first order in $g_s$ and matching them with perturbations found from the formula \eqref{bulkb}. On the contrary, the emission found here are completely new and at the moment there is no known SUGRA  solution in literature written in terms of arbitrary harmonic functions that can accommodate for the microstates found in string theory.  We conjecture that such SUGRA solutions should exist and we give some arguments (section 5) supporting this thesis. It would be relevant to write explicitly such solution and we plan to address this issue in a future work.

Two important points were left open in \cite{Bianchi:2016bgx} and are not closed here. 
First of all,  since the number of disk with different boundaries grows with the product of the charges like $Q_1 Q_2 Q_3 Q_4$ and the  SUGRA solution is dual to this configuration, we expect that there should be some way to compute the entropy directly from gravity, on the same line of previous successes for the D1D5 case \cite{Rychkov:2005ji,Krishnan:2015vha}, without dealing with the  complicated $\mathcal{N}=4$ quantum mechanics on the branes worldvolume\footnote{See for instance   \cite{Chowdhury:2015gbk} for BPS state counting in a  pure D-brane configuration, even though the calculation is performed in the limit of small charges, that is the opposite of the SUGRA limit.}. Second, to obtain the fuzzball of a black hole  it is necessary to show that regular, horizonless and asymptotically flat solutions exist, and this means finding the explicit form of the harmonic function, 
similarly to what has been done in \cite{Lunin:2015hma} for $AdS_2 \times S_2$  asymptotics. We plan to also come back on this issues in a future work.

\section{One Boundary Amplitude}

In this section we compute the disk amplitude for the  emission of one closed NSNS field from a single stack of  Dp-branes at angles\footnote{Since there is only one stack, the angle is between the branes and  a fixed reference frame. This angle is actually meaningless in an uncompactified geometry, where it can be reabsorbed by a change of coordinates, while is meaningful and not completely arbitrary in a compact space. }, in the  weak coupling regime. By using \eqref{bulkb} we will be able to extract the corresponding gravitational background generated in the SUGRA limit.

This calculation has already been performed in \cite{Bianchi:2016bgx} for perpendicular intersecting branes. The procedure is identical for tilted branes, the only difference is the expression of the reflection matrix $R$; therefore in this section we simply state the result of the amplitude and and setup notation for the next sections.
The relevant vertex operators are
  
\bea
W_{NSNS}(z, \bar z) &=&   (E R)_{MN}  \,e^{-\varphi} \psi^{M} e^{ik X} (z)  \, e^{-\varphi} \psi^{N} e^{ik  R X} (\bar z) \nn\\
V_{\xi(\phi)}({x}_a) &=& 
 \sum_{n=0}^\infty  \xi_{i_1\ldots i_n}  \,   \partial X^{i_1}  (x_1) \, 
\prod_{a=2}^n\, \int_{-\infty}^\infty \,    {d{x}_a\over 2\pi  }  \,   \partial X^{i_a}  ({x}_a)
\eea

where $E=h+b$ is the polarization, that encompasses both the graviton and the B-field,  $R$ is the reflection matrix  encoding the boundary conditions of the open strings, $\varphi$ is a scalar field that bosonize the superghost, and $\psi$ and $X$ are the worldsheet fermions and bosons respectively. While the first vertex describe closed strings, the second vertex stands for an arbitrary number of untwisted (with  endpoints on the same stack of branes) open string scalar fields  that can be inserted on the boundary, with some polarization $\xi_{i_1\ldots i_n}$. Remarkably, the insertion of these fields  translates into the full multipole expansion of a generic harmonic function appearing in the SUGRA fields.

The amplitude to evaluate is:

\begin{equation}
\mathcal{A}_{NS{-}NS,\xi(\phi)}=  \langle c(z) c(\bar z) c(z_1) \rangle  \left\langle  W_{NS{-}NS}(z,\bar z)      V_{\xi(\phi)}   \right\rangle
\end{equation}

Left and right string sectors are related by:

\begin{equation}
X_{right}^{M}=R^{M}_{N} X^{N}_{left} \;\;\;\; \psi_{right}^{M}=R^{M}_{N} \psi_{left}^{N} \;\;\;\;  \varphi_{right}=\varphi_{left}
\end{equation}

The ten dimensional reflection matrix enforce the boundary conditions for branes at angle and therefore contains a non trivial internal part in $T^6$:

\begin{equation}
R^{M}_{\,\,\,\,\, N}= \begin{pmatrix} 1 & 0 & 0 & 0 &0\\
0 & -\mathbb{1}_{3 \times 3} & 0 & 0 & 0 \\ 
 0 & 0 & R(\theta_1) & 0 &0\\
  0 & 0 & 0 & R(\theta_2)  &0 \\
   0 & 0 & 0 & 0& R(\theta_3)  
 \end{pmatrix}  
 \label{refl}
\end{equation}

where we have introduced the two dimensional matrices derived in the appendix:

\begin{equation}
R(\theta_i) = \begin{pmatrix} cos(2\theta_i) & sen(2\theta_i)\\
sen(2\theta_i) & -cos(2\theta_i) \end{pmatrix} 
\end{equation}

The final result is:

\bea
{\mathcal{A}_{NSNS, \xi(\phi) } =  \,  ({E}R)  \,{  \xi(  k)  } }  
\eea

where $ER=E_{M N} R^{N}_{\,\,\,\,\, P} \eta^{PM}$ and $\xi(k)=\sum_{n=0}^\infty \xi_{i_1\ldots i_n } k^{i_1} \ldots k^{i_n}$. 

The emitted field is the graviton, therefore using \eqref{bulkb} and Fourier transforming (setting here and in the following $ \xi(  k)$ to a constant for simplicity) with 
\bea
\int \frac{ d^3k}{(2 \pi)^3} \frac{1}{ k^2} e^{ikx}= \frac{1}{4 \pi} \frac{1}{r} 
\eea
 one finds that the deviation from flat space is given by the schematic form:

\bea
h_{MN}=\frac{(\eta R)_{MN}}{r}
\label{hm}
\eea

where all the inessential constants have been absorbed into $ \xi(  k)$. Notice that the result is proportional to $g_s$, indeed the factor $g_s^2$ coming from the massless closed string propagator in \eqref{bulkb} partially cancels with the $g_s^{-1}$ associated to the disk (open strings don't carry additional $g_s$ factors). 

This result can be directly compared to the SUGRA metric 

\bea
ds^2=L^{-\ft{1}{2}}(-dt^2+\sum_{I=1}^3 d\tilde{y}_I^2)+L^{\ft{1}{2}}(\sum_{I=1}^3 dy_I^2+\sum_{i=1}^3 dx_i^2)
\eea
of D3 brane with tilted internal axes. Indeed  (focusing on the first torus) performing the rotation in the internal tori as 
\bea
\tilde{y}'_1=\cos(\theta) \tilde{y}_1-\sin(\theta) y_1 \nn \\ y'_1=\sin(\theta) \tilde{y}_1+\cos(\theta) y_1
\eea
 and expanding at the leading order in $g_s$ with the ansatz $L=1+\ft{ g_s}{r}$ one finds:

\bea
h_{MN}dx^{M}dx^{N} = \ft{g_s}{2} (dt^2+\sum_{i=1}^3 dx_i^2+ \cos(2\theta) dy_1^2 +2 \sin(2\theta)dy_1 d\tilde y_1-\cos(2\theta)d\tilde y_1^2 ) \nn \\
\eea

having omitted a similar contribution from the other two internal torii. This expression matches what appears in \eqref{hm} after reabsorbing the overall constants.

\section{Vertex Operators for Branes at Angle}

Boundary conditions imposed by intersecting branes at angle force the open strings stretched between the branes to have non integer mode expansions, therefore leading to vertex operators   involving angle-dependent bosonic and fermionic twist field. The T-dual picture is given by open strings ending on D-branes with magnetic fields switched on along the world-volume. In the following we will construct a supersymmetric system of 4 D3 branes  at angles and then we will identify the  vertex operators corresponding to each pair of branes.

 Every Dp-brane imposes some restrictions on the spinors $\epsilon_{L,R}$ parametrizing the SUSY transformations generated by $\epsilon_L Q_L+\epsilon_R Q_R$, where $Q_{L,R}$ are the supercharges (with the same 10d chirality) of $IIB$ superstring theory. Solutions to these constraints count the number of unbroken supercharges in the system and when angles are taken into account one finds that if the branes are related by $SU(3)$ rotations some supersymmetry is preserved \cite{Berkooz:1996km}. In particular, we require that at least $\mathcal{N}=1$ is preserved between every couples of branes, leading to a condition on the (relative) angles:

\bea
\theta_1+\theta_2+\theta_3=0 \qquad mod  \,\, 2\pi
\label{angle}
\eea

where $\theta_I$ is the angle between two branes in the $\lbrace  y_I,\tilde{y}_I  \rbrace$ torus. To preserve $\mathcal{N}=2$,  \eqref{angle} must be true and  an angle must be zero, and to have $\mathcal{N}=4$ all the angles need to be zero. 

\begin{figure}
\begin{center}
\includegraphics[width=8cm]{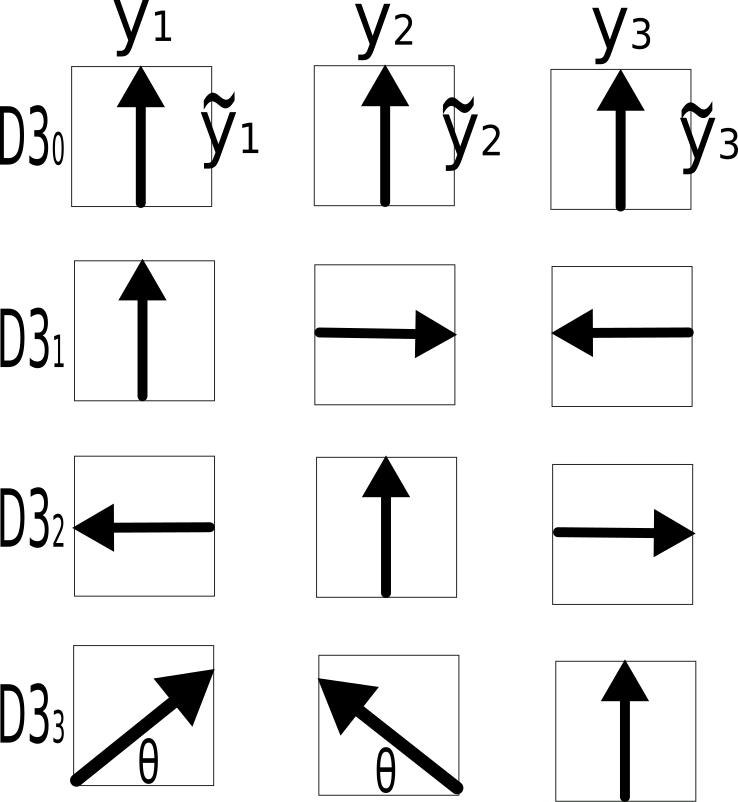} 
\end{center}
\caption{D3-brane configuration in the internal tori.}
\label{fig4}
\end{figure}

One can verify that the system of Fig.\eqref{fig4} satisfies \eqref{angle} for all the six possible pairs of branes.  The convention is that a positive angle is taken counterclockwise from the horizontal $y$ direction, so for instance a right arrow stands for $\theta=0$, a left arrow for $\theta=\pi$ and up arrow for $\theta=\pi/2$.

To be concrete, we construct the vertex operators for the fermionic open string stretched between the (ordered) pair $D3_0D3_3$  (for a more general discussion on vertex operators at angle see \cite{Cvetic:2006iz}). In the canonical $-\ft{1}{2}$ superghost picture the vertex operator is:\footnote{ The subscript $\mu$ in $V_{\mu}$ is a symbol for the polarization, not an index!}

   \begin{equation}
 V_{\mu}^{\lbrace 03 \rbrace}(z_n)=e^{-\ft{\varphi}{2}} \mu^{\beta}_{\lbrace 03 \rbrace}  S_{\beta} (\sigma^{(1)\dagger}_{1/2-\theta} e^{  i  \theta\varphi_1}) (\sigma^{(2)}_{1/2-\theta}e^{ - i \theta \varphi_2})(e^{ - i \varphi_3/2})e^{ik_{n}X}
\end{equation}

where $S_{\beta}$ is an $SO(1,3)$ spin field (see the appendix for notation and conventions) and $\mu^{\beta}$ is the fermionic polarization with Chan Paton indices  not shown explicitly. In particular every polarization $\mu^\beta$ is actually a Chan-Paton matrix, for instance $(\mu_{\lbrace 03 \rbrace}^{a \, b})^\beta$ refers to the open string stretched between the brane $a$ in the stack 0 and the brane $b$ in the stack 3 (the order is important). 

A bosonic twist $\sigma^{(I)}_{\xi}$ field must be inserted for every $I$-torus ($I=1,2,3$) in which the branes are not parallel. The twist field is angle  dependent and the angle $\xi$ is computed by rotating the first brane, the $D3_0$, to the second brane, with a dagger if the rotation is clockwise and without dagger if it is counterclockwise. Of course $\sigma_{\theta}^{\dagger}=\sigma_{1-\theta}$.  Similarly, one must add a fermionic twist $e^{i (\xi-\ft{1}{2}) \varphi_I}$ for every torus. As a check one can verify that the conformal dimension\footnote{The conformal dimension of a bosonic twist field is given by $h(\sigma_{\theta})=\ft{1}{2} \theta (1-\theta)$. Other useful formulas are $h(e^{q\varphi})= -\frac{1}{2} q^2-q$ and  $h(e^{i\lambda\varphi})= \frac{1}{2} \lambda^2$
.} of the vertex operator is equal to 1 and that the vertex is consistent with the SUSY condition \eqref{angle}. Indeed the action of of the $IIB$ positive chirality supercharge $Q^{(+++++)}=\oint dz e^{+\frac{i}{2}(\varphi_4+\varphi_5+\varphi_1+\varphi_2+\varphi_3)} e^{-\varphi /2}$  must be well defined, implying  that  no non-integer powers of $z$ can appear in the commutator $
[\epsilon Q ,V_{\mu}^{(-1/2)}]$, which gives the susy-related bosonic vertex operator.

The opposite chirality fermion, the one stretched from $D3_3$ to $D3_0$ is recovered by doing the conjugation of the previous vertex:

\begin{equation}
 V_{\bar \mu}^{\lbrace 30 \rbrace}(z_n)=e^{-\ft{\varphi}{2}} \bar \mu^{\dot{\beta}}_{\lbrace 30 \rbrace}  C_{\dot{\beta}} (\sigma^{(1)}_{1/2-\theta} e^{ - i  \theta\varphi_1}) (\sigma^{(2)\dagger}_{1/2-\theta}e^{  i \theta \varphi_2})(e^{  i \varphi_3/2})e^{ik_{n}X}
\end{equation}

\section{Two Boundary Amplitude}

 In this section we evaluate the disk amplitude for the scattering of one closed NSNS fields with two open strings stretched between two stacks of intersecting tilted D3 branes. This case already illustrates the crucial features that will be present in the more physically interesting four boundary computation. Similar calculation without non trivial angles have already been performed for $D1D5$ and $D3D3'$ perpendicularly intersecting system in \cite{Giusto:2009qq,Giusto:2011fy,Bianchi:2016bgx}.  The system considered is a non threshold bound state of D-branes, meaning that the solution is not in a naive BPS superposition and individual branes cannot be freely separated (Higgs branch). From the microscopical point of view, this is signaled by the presence of a non zero open string condensate, or in other words of a non zero  v.e.v. for the fields associated to open strings stretched between different branes. 

The string amplitude to be computed is

\begin{equation}
\mathcal{A}_{\mu\bar{\mu} NSNS}=\int \frac{d^4z}{V_{CKV}}     \langle V^{(-1/2)}_{\mu}(z_1) V^{(-1/2)}_{\bar{\mu}}(z_2)W^{(0,-1)}_{NSNS}(z_3,z_4) \rangle
\label{01closed}
\end{equation}

in which $V_{CKV}$ is the conformal Killing volume, $z_1$ and $z_2$ are real, while $z_3=\bar{z}_4$ is complex.

\begin{table}[h]
\begin{center}
\begin{tabular}{|c|c|c|c|c|c|c|c|c|c|c|}
\hline
        Brane     &  t       &$x_1$   & $x_2$  &$x_3$  & $y_1$ &$\tilde{y}_1 $  &$y_2$  &$\tilde{y}_2$   & $y_3$ &$\tilde{y}_3$          \\
\hline
 $ D3 $      & $-$      &$.$     & $.$   &$.$    & $.$   &$-$    &$.$    &$-$    & $.$   &$-$        \\
$ D3'$           &  $-$      &$.$    & $.$   &$.$    & $\theta$  &$ $    &$\pi-\theta$   &$$    & $.$   &$-$  \\
\hline
\end{tabular}
\caption{D3-brane configuration: Neumann and Dirichlet directions are lines and dots respectively. The angle is taken with respect to the non tilded y coordinate.  }
\end{center}
\end{table}

The brane configuration can be taken to be the $D3_0D3_3$ as seen before. Here we rewrite the vertex operators needed for the calculation, with a superscript indicating the superghost charge:

\begin{equation}
 V^{(-1/2)}_{\mu}(z_1)=\mu_{\alpha}  S^{\alpha} e^{-\varphi /2} (\sigma^{(1)\dagger}_{1/2 -\theta}e^{  i \theta\varphi_1}) (\sigma^{(2)}_{1/2 -\theta}e^{ - i \theta\varphi_2})(e^{ - i \varphi_3/2})e^{ik_{1}X}
\end{equation}

\begin{equation}
 V^{(-1/2)}_{\bar{\mu}}(z_2)=\bar{\mu}_{\dot{\alpha}}  C^{\dot{\alpha}}e^{-\varphi /2}  (\sigma^{(1)}_{1/2 -\theta}e^{  -i \theta\varphi_1}) (\sigma^{(2)\dagger}_{1/2 -\theta}e^{  i \theta\varphi_2})(e^{  i \varphi_3/2})e^{ik_{2}X}
\end{equation}

\begin{equation}
W_{NSNS}^{(0,-1)}(z_3,z_4)=(ER)_{MN} (\partial X^{M}-ik  \psi \psi^{M})e^{ikX}{(z_3)}e^{-\varphi} \psi^{N} e^{ikR X}{(z_4)} 
\label{closedw}
\end{equation}

The reflection matrix can be choosen to be the one associated to the $D3_0$ or the one associated to the $D3_3$, the final result will be independent of this choice.
The spinorial structure of the condensate $v$ is :

\begin{equation}
 \mu^{\alpha} \bar{\mu}^{\dot{\beta}} \;  = \; (\bar{\sigma}^{\mu})^{\dot{\beta} \alpha}
 v_\mu\end{equation}
 
The open string condensate microscopically encodes the different microstates of the black hole geometries. Notice that the  condensate $v$ can be linked to the profile function $f(v)$ appearing as a displacement of the center of the $D1D5$ fuzzball harmonic functions or alternatively as the profile of the oscillation of the string in the U-dual F1-P (fundamental string with momentum) system \cite{Lunin:2001fv}\nocite{Lunin:2001jy,Lunin:2002bj,Lunin:2002iz}-\cite{Kanitscheider:2007wq}. More precisely the condensate is linked to the derivative of $f(v)$ \cite{Giusto:2009qq}, therefore in some sense it gives the dynamic part of the displacement, whether the fixed part can be encoded in the untwisted scalars, as we have already said. While here the condensate is an arbitrary real number, in a proper treatment based on the microscopic theory on the world-volume of the D-branes it should be possible to quantize it, matching all the possible open string configuration to the respective condensate.

In the following we focus on constants open string fields, therefore we can consistently take all the open strings momenta to be vanishing. As we will see, care is needed in taking the zero momentum limit, indeed the amplitude is irreducible only if the disk diagram cannot factorize into two or more disk diagrams connected by propagating massless open fields. If the diagram factorize, the amplitude has massless open string poles at zero momentum exchanged; the associated worldsheet integral is divergent and there is no corresponding SUGRA emission. The closed string momentum satisfies  $k^2=0$ and $k^{M}E_{MN}=0$, moreover we concentrate on a purely spatial momentum $k^M=\lbrace k^0,k^i,k^{int} \rbrace=\lbrace 0,k^i,0\rbrace$,  analytically continuing the momentum to complex values in order to be consistent with the mass-shell condition.

Here we report directly the result of the computation, the reader interested in the details can found the explicit derivation in the appendix. The non compact part and the $(M,N)=(\bar{3},3)$ component have a finite amplitude given by:

\begin{equation}
\mathcal{A}^{NC}_{\mu\bar{\mu} NSNS} = \frac{\pi }{\sqrt{2}}   (ER)(kv) 
 \end{equation}

\begin{equation}
\mathcal{A}^{3\bar3}_{\mu\bar{\mu} NSNS} = \frac{-\pi }{\sqrt{2}}    (ER)_{[3 \bar{3}]} (k v) 
 \end{equation}
 
 with $vk=v^i k^i$, and $(\bar{3},3)$ being the complex directions defined from the real ones $(\tilde{3},3)$. On the contrary the tilted directions give rise to divergent or vanishing amplitudes, depending on the symmetry property of the polarizations.
 The divergent part require some clarification: it implies that there is no associated SUGRA emission for that choice of the condensate. What's happening can be understood clearly  from the $SO(1,5)$ picture of the directions $\lbrace t, \, x_1, \,x_2, \, x_3, \, y_3, \, \tilde{y}_3 \rbrace$: the fermionic condensate $ \mu^A \bar{\mu}^B$, where $A,B=1, \dots ,4$, factorizes as $\bf{4} \times 4 = 6 + 10$. The $\bf{6}$, the anti-symmetric part of the product, is simply the vectorial representation, corresponding to the $\Gamma^{\hat{M}}_{[AB]}$  ($\hat{M}=1,\dots ,6$) appearing in the fermionic kinetic term in the effective lagrangian, therefore giving a factorization channel for the $NSNS$-fermion-fermion diagram into two diagrams linked by a massless scalar field $\phi_{[AB]}$. To get a non factorizable diagram, one must restrict to the symmetric part of the product for which no fields with suitable indices are present.

A simple way to choose the symmetric part is to restrict to $S^A=S^B=e^{\frac{i}{2}(-\varphi_3-\varphi_4-\varphi_5)}$. The vertices to employ are then:
 
 \begin{equation}
 V^{(-1/2)}_{\mu}(z_1)=\mu_{A}  S^{A} e^{-\varphi /2} (\sigma^{(1)\dagger}_{1/2 -\theta}e^{  i \theta\varphi_1}) (\sigma^{(2)}_{1/2 -\theta}e^{ - i \theta\varphi_2})e^{ik_{1}X}
\end{equation}

\begin{equation}
 V^{(-1/2)}_{\bar{\mu}}(z_2)=\bar{\mu}_{B}  S^{B}e^{-\varphi /2}  (\sigma^{(1)}_{1/2 -\theta}e^{  -i \theta\varphi_1}) (\sigma^{(2)\dagger}_{1/2 -\theta}e^{  i \theta\varphi_2})e^{ik_{2}X}
\end{equation}

resulting in the $SO(1,5)$ amplitude:

\bea
\mathcal{A}^{SO(1,5)}_{\mu\bar{\mu} NSNS}  =  \ft{1}{3!}\, (E R)_{\hat{M}\hat{N}} k_{\hat{P}}  \, v^{\hat{M}\hat{N}\hat{P}}
\label{3c}
 \eea
 
 having defined a three index condensate contracting $\Gamma^{\hat{M}\hat{N}\hat{P}}_{(AB)}$ with $\mu^{(A} \bar{\mu}^{B)}$ and having used hatted $SO(1,5)$ indices.
 Notice that in the final results the angles disappeared altogether. In order to see a non trivial angular dependence one must break the $SO(1,5)$ symmetry by tilting even the third torus, therefore dealing with three point correlation functions of twist fields. 
 Such correlators will be present in the next section, resulting in an angular dependence of the amplitudes.

\section{Four Boundary Amplitude}

Finally we consider the $1/8$ BPS configuration of the 4 stack of branes oriented like in Fig.\eqref{fig4}. The string amplitudes correspond to a disk with four different boundaries and polarizations $\mu$ chosen in a cyclic order starting and ending on  $D3_0$. The associated condensate is complex, meaning that one can run the cycle in two different directions using $\mu$ or $\bar{\mu}$ respectively.

\begin{table}[h]
\begin{center}
\begin{tabular}{|c|c|c|c|c|c|c|c|c|c|c|}
\hline
        Brane     &  t       &$x_1$   & $x_2$  &$x_3$  & $y_1$ &$\tilde{y}_1 $  &$y_2$  &$\tilde{y}_2$   & $y_3$ &$\tilde{y}_3$          \\
\hline
 $ D3_0 $      & $-$      &$.$     & $.$   &$.$    & $.$   &$-$    &$.$    &$-$    & $.$   &$-$        \\
  $ D3_1 $      & $-$      &$.$     & $.$   &$.$    & $.$   &$-$    &$\rightarrow$    &$.$    & $\leftarrow$   &$.$        \\
   $ D3_2 $      & $-$      &$.$     & $.$   &$.$    & $\leftarrow$   &$.$    &$.$    &$-$    & $\rightarrow$   &$.$        \\
$ D3_3$           &  $-$      &$.$    & $.$   &$.$    & $\theta$  &$ $    &$\pi-\theta$   &$$    & $.$   &$-$  \\
\hline
\end{tabular}
\caption{D3-brane configuration: Neumann and Dirichlet directions are lines and dots respectively. The arrow distinguishes between the $\theta=0$ (right arrow) and $\theta=\pi$ (left arrow) case. The angle is taken with respect to the non tilded y coordinate.  }
\end{center}
\end{table}

 The amplitude to be computed is:

\begin{equation}
\mathcal{A}_{\mu^4 NSNS}=\int \frac{d^4z}{V_{CKV}}     \langle V^{(-1/2)}_{\mu}(z_1) V^{(-1/2)}_{\mu}(z_2)V^{(-1/2)}_{\mu}(z_3)V^{(-1/2)}_{\mu}(z_4) W^{(0,0)}_{NSNS}(z_5,z_6) \rangle
\end{equation}

with vertex operators constructed as previously explained:

\begin{equation}
 V_{\mu}^{\lbrace 01 \rbrace}(z_1)=e^{-\ft{\varphi}{2}} \mu^{\alpha}_{\lbrace 01 \rbrace}  S_{\alpha} (e^{  -i \varphi_1/2}) (\sigma^{(2)\dagger}_{1/2})(\sigma^{(3)}_{1/2})e^{ik_{1}X}
\end{equation}

\begin{equation}
 V_{\mu}^{\lbrace 12 \rbrace}(z_2)=e^{-\ft{\varphi}{2}} \mu^{\beta}_{\lbrace 12 \rbrace}  S_{\beta} (\sigma^{(1)}_{1/2}) (\sigma^{(2)}_{1/2})(e^{  -i \varphi_3/2})e^{ik_{2}X}
\end{equation}

\begin{equation}
 V_{\mu}^{\lbrace 23 \rbrace}(z_3)=e^{-\ft{\varphi}{2}} \mu^{\dot{\alpha}}_{\lbrace 23 \rbrace}  C_{\dot{\alpha}} (\sigma^{(1)}_{\theta} e^{  i \varphi_1 (\theta+1/2)}) (\sigma^{(2)}_{1/2-\theta}e^{  -i \theta \varphi_2})(\sigma^{(3)}_{1/2})e^{ik_{3}X}
\end{equation}

\begin{equation}
 V_{\mu}^{\lbrace 30 \rbrace}(z_4)=e^{-\ft{\varphi}{2}} \mu^{\dot{\beta}}_{\lbrace 30 \rbrace}  C_{\dot{\beta}} (\sigma^{(1)}_{1/2-\theta} e^{ - i  \theta\varphi_1}) (\sigma^{(2)\dagger}_{1/2-\theta}e^{  i \theta \varphi_2})(e^{  i \varphi_3/2})e^{ik_{4}X}
\end{equation}

\bea
W^{(0,0)}(z_5,z_6) = c_{\rm NS} \, ({E}R)_{MN}  (\partial  X^{M}-i \, k  \psi \, \psi^{M})  e^{\ii k X}(z_5)
(\partial  X^{N}-i \, k R  \psi \, \psi^{N}) e^{\ii k R X}(z_6) \nn \\
\eea

It's convenient to take the choice $  \, \mu_{\lbrace 01 \rbrace}^{(\alpha} \, \mu_{\lbrace 12 \rbrace}^{\beta)}\,  \bar\mu_{\lbrace 23 \rbrace}^{(\dot\alpha}\,  \bar\mu_{\lbrace 30 \rbrace}^{\dot\beta)}  $for the condensate. This configuration  lives in the $(\textbf{3}_L,\textbf{3}_R)$, that can be seen as a symmetric traceless tensor $v^{ij}$, and leads to non factorizable diagrams.

In particular let's fix:
\bea
   S_{\alpha}= S_{\beta}=e^{  {\ii\over 2}(\varphi_4 +\varphi_5 ) }   \qquad   C_{\dot \alpha}=C_{\dot \beta}=e^{-{\ii\over 2}(\varphi_4-\varphi_5) }    
\eea

therefore the only way to saturate the $+2$ superghost charge in the $\lbrace  5,\bar{5} \rbrace$  torus is taking the term with four fermions in the closed string vertex. We split the calculation in non-compact and  compact directions. We start by computing the correlators common to both cases:
   
 \bea
 && \langle c(z_1)\,c(z_2)\,c(z_4)\rangle =  z_{12}\, z_{24}\, z_{41}  \nn \\
 && \left\langle \prod_{j=1}^4 e^{-\varphi/2}(z_j) \right\rangle =
\prod_{i<j}  z_{ij}^{-1/4 } \nn\\ 
\label{1corr}
\eea

The twist fields correlator in the third torus is a two point function, while in the other torus we have a three and a four point function. The latter correlators in principle receive a  classical contribution proportional to the area of the polygon formed by the intersecting branes, but it can be neglected in the SUGRA limit (\cite{Cvetic:2006iz,Cvetic:2007ku}). The explicit expressions for the quantum part are:

\bea
 &&   
\left\langle   \sigma^{(3)}_{1/2} (z_1)    \sigma^{(3)}_{1/2} (z_3)  \right\rangle  = z_{13} ^{-\ft{1}{4} }    \nn\\
 &&   
\left\langle    \sigma^{(1)}_{1/2} (z_2)    \sigma^{(1)}_{\theta} (z_3) \sigma^{(1)}_{1/2-\theta} (z_4)   \right\rangle  =(4 \pi \Gamma_{ \lbrace\ft{1}{2}, \theta, \ft{1}{2}-\theta \rbrace})^{\ft{1}{4}} z_{23}^{-\ft{\theta}{2} }   z_{24}^{+\ft{\theta}{2}-\ft{1}{4} }   z_{34}^{-\ft{\theta}{2}+\theta^2  }   \nn\\
&&   
\left\langle    \sigma^{(2)}_{1/2} (z_1)    \sigma^{(2)}_{1/2} (z_2)    \sigma^{(2)}_{1/2-\theta} (z_3)  \sigma^{(2)}_{1/2+\theta} (z_4)   \right\rangle  = z_{12}^{-\ft{1}{4} } z_{34}^{-\ft{1}{4}+\theta^2 } \left( \frac{ z_{13} z_{24}}{ z_{14} z_{23}} \right)^{\ft{1}{4}} \mathcal{I}^{-\ft{1}{2}}[w]
\label{twist}
 \eea
 
 where we have used
 
 \bea
  \left\langle   \sigma_{\alpha} (z_1)\sigma_{\beta} (z_2)\sigma_{\gamma} (z_3)  \right\rangle  =(4 \pi \Gamma_{ \lbrace \alpha, \beta, \gamma \rbrace})^{\ft{1}{4}} z_{12}^{-\alpha \beta }  z_{13}^{-\alpha \gamma} z_{23}^{-\beta \gamma }    \nn\\
 \eea
 
\bea
\Gamma_{\alpha, \beta, \gamma}=\frac{\Gamma(1-\alpha)\Gamma(1-\beta)\Gamma(1-\gamma)}{\Gamma(\alpha)\Gamma(\beta)\Gamma(\gamma)}
\eea 

\bea
\left\langle    \sigma_{1-a} (z_1)  \sigma_{a} (z_2) \sigma_{1-b} (z_3)  \sigma_{b} (z_4)    \right\rangle  = z_{12}^{-a(1-a) } z_{34}^{-b(1-b) } \left( \frac{ z_{13} z_{24}}{ z_{14} z_{23}} \right)^{\ft{1}{2}(a+b)-ab} \mathcal{I}^{-\ft{1}{2}}(w)
\eea

\bea
\mathcal{I}[w]=\frac{1}{2\pi} (B_1(a,b) G_2(w) H_1(1-w)+B_2(a,b) G_1(w) H_2(1-w))
\eea

\bea
B_1(a,b)= \frac{\Gamma(a)\Gamma(1-b)}{\Gamma(1+a-b)} \qquad B_2(a,b)= \frac{\Gamma(b)\Gamma(1-a)}{\Gamma(1+b-a)}
\eea

\bea
G_1(w)= {}_2F_1[a,1-b; 1; w] \qquad G_2(w)= {}_2F_1[1-a,b; 1; w]
\eea

\bea
H_1(w)= {}_2F_1[a,1-b; 1+a-b; w] \qquad H_2(w)= {}_2F_1[1-a,b; 1-a+b; w]
\eea

\bea
w=\frac{ z_{12} z_{34}}{ z_{13} z_{24}}=1-x
\eea 

 where the last equality is valid after the gauge fixing 
 
  \be
  z_1=-\infty \qquad z_2=0\qquad z_3=x \qquad z_4=1 \qquad z_5=z \qquad z_6=\bar{z}
  \label{fix}
  \ee
\subsection{Fermions in Extended Space-time  Dimensions}

Now we specialize to the amplitude with  the choice $(M,N)=(4,\bar 4)$: all the worldsheet fermions $\psi$ are in the extended space-time directions.

\bea
&& 
\left\langle   e^{  -i \varphi_1/2}(z_1)  e^{  i \varphi_1(\theta+1/2)}(z_3)  e^{  -i \theta \varphi_1}(z_4)  \right\rangle 
 =  z_{13}^{-{\theta\over 2}-\ft{1}{4} } z_{14}^{{\theta\over 2}}  z_{34}^{-\theta^2-\ft{\theta}{2} }  \nn\\
 && 
\left\langle     e^{  -i \theta \varphi_2}(z_3)  e^{  i \theta \varphi_2}(z_4)  \right\rangle 
 =   z_{34}^{-\theta^2}  \nn\\
 && 
\left\langle     e^{  -i  \varphi_3/2}(z_2)  e^{  i  \varphi_3/2}(z_4)    \right\rangle 
 =   z_{24}^{-\ft{1}{4}} 
 \label{4bound33}
\eea 
 
 The correlator of the four spin fields and four fermions fields after bosonization is:

 \bea
&& \left \langle e^{  {\ii\over 2}(\varphi_4 +\varphi_5 ) }_{(z_1)} e^{  {\ii\over 2}(\varphi_4 +\varphi_5 ) }_{(z_2)} e^{ - {\ii\over 2}(\varphi_4 -\varphi_5 ) }_{(z_3)} e^{  -{\ii\over 2}(\varphi_4 -\varphi_5 ) }_{(z_4)} e^{  \ii(\varphi_4 -\varphi_5 ) } _{(z_5)}    e^{  \ii(-\varphi_4 -\varphi_5 ) }_{(z_6)}  \right\rangle   =   \ft{1}{4}  z_{12}^{\ft{1}{2}}   z_{16}^{-1} z_{26}^{-1}  z_{34}^{\ft{1}{2}} z_{35}^{-1} z_{45}^{-1}  \nn \\
\eea

Using the gauge fixing \eqref{fix},  the remaining  worldsheet  triple integral is\footnote{Notice that by putting $\theta=0$ in one recover the case of  perpendicular branes, involving the correlator of 4 $\mathbb{Z}_2$ twist fields \cite{Bianchi:2016bgx}. In particular $\mathcal{I}^{-\ft{1}{2}}[1-x]_{\theta=0}=\left( {}_2F_1[\ft{1}{2},\ft{1}{2}; 1; x]{}_2F_1[\ft{1}{2},\ft{1}{2}; 1; 1-x] \right)^{-1/2} $.}:

 \begin{equation}
I_4=\int_{0}^{1} dx    (-x)^{-\ft{1}{2}-\ft{\theta}{2}} (x-1)^{-\theta} \mathcal{I}^{-\ft{1}{2}}[1-x]\int_{\mathbb{C}^+} \frac{d^2z }{  \bar z (1- x) (x- z)}
\end{equation}

that can be reduced to a single integration by computing the integral in the upper half of the complex plane (for instance using $z=r e^{i \alpha}$ and performing the r integration first):

\begin{equation}
\int_{\mathbb{C}^+} \frac{d^2z }{  \bar z (1- x) (x- z)}=- \frac{\pi log(x)}{(x-1)}
\end{equation}

The remaining integral must be performed numerically and is a finite overall constant that can be reabsorbed into the open string condensate. In the particular case of $\theta=0$ the integral is purely imaginary, but for a generic angle the result has a real part too. 

Recombining all the results obtained so far one obtains the amplitude:

 \be
\mathcal{A}^{NC}_{\mu^4 NSNS}=     ({E}R)_{4\bar 4} \,   k_{i} k_{j}  \,    v^{ij}(\theta) 
\ee   

where all the constants, group theory factors and the Chan-Paton trace over the polarizations have been reabsorbed into the symmetric traceless condensate $v^{ij}(\theta)$. The real and imaginary components of the amplitude contribute respectively to the symmetric and antisymmetric parts of $(ER)_{M \bar{M}}$, therefore for a generic angle and generic complex condensate the system will emit both the graviton and the B-field\footnote{It's nevertheless possible to select the complex open string condensate in such way that only the graviton is emitted.}.  The associated  harmonic is given by $\frac{3x_j x_i-\delta_{ij}r^{2}}{ r^{5}}$, having used 

\bea \int \frac{ d^3k}{(2 \pi)^3} \frac{-i k_j k_i}{ |k|^2} e^{ikx}= \frac{i}{4 \pi} \frac{3x_j x_i-\delta_{ij}|x|^{2}}{ |x|^{5}}.
\eea


\subsection{Fermions in Compact Dimensions}

Now we focus on the choice $(M,N)=(J,\bar{J})$ with $J=1,2,3$.
We have already computed the correlators \eqref{1corr} and \eqref{twist}. The correlator of spin fields and fermions is the same for all $J$:

 \bea
&& \left \langle S_{(\alpha}(z_1) S_{\beta)}(z_2) C_{(\dot \alpha}(z_3) C_{\dot\beta)}(z_4) \psi^{ \mu} (z_5)     \psi^{ \nu} (z_6)  \right\rangle   =  
  { (z_{12} z_{34} )^{1/2} z_{56} \over  
2 \prod_{i=1}^4 (z_{i5} z_{i6})^{1/2}  }  \sigma^{(\mu}_{\alpha \dot{\alpha}  }  \sigma^{\nu)}_{\beta \dot \beta} 
\eea

In the following we list the remaining correlators and the final gauge fixed integrals.

\begin{itemize}
\item  $(M,N)=(1,\bar{1})$

\bea
&& 
\left\langle   e^{  -i \varphi_1/2}_{(z_1)}  e^{  i \varphi_1(\theta+1/2)}_{(z_3)} e^{  -i \theta \varphi_1}_{(z_4)} \psi^1 _{(z_5)}     \psi^{\bar 1}_{(z_6)}  \right\rangle 
 = \ft{1}{2} z_{13}^{-{\theta\over 2}-\ft{1}{4} } z_{14}^{{\theta\over 2}}  z_{34}^{-\theta^2-\ft{\theta}{2} } z_{15}^{-\ft{1}{2}}  z_{16}^{\ft{1}{2}} z_{35}^{{\theta\over 2}+\ft{1}{2}}  z_{46}^{-{\theta\over 2}-\ft{1}{2}}   z_{45}^{-\theta}  z_{46}^{\theta}  z_{56}^{-1} \nn\\
 && 
\left\langle     e^{  -i \theta \varphi_2}(z_3)  e^{  i \theta \varphi_2}(z_4)  \right\rangle 
 =   z_{34}^{-\theta^2}  \nn\\
 && 
\left\langle     e^{  -i  \varphi_3/2}(z_2)  e^{  i  \varphi_3/2}(z_4)  \right\rangle 
 =   z_{24}^{-\ft{1}{4}}  
\eea 
 \begin{equation}
I_1=\int_{0}^{1} dx    (-x)^{-\ft{1}{2}-\ft{\theta}{2}} (x-1)^{-\theta} \mathcal{I}^{-\ft{1}{2}}[1-x]\int_{\mathbb{C}^+} \frac{d^2z \, (x- z)^{\theta}(x- \bar{z})^{\theta-1}}{  |z| (1- z)^{\theta+1/2}  (1- \bar{z})^{-\theta+1/2}}
\end{equation}


\item  $(M,N)=(2,\bar{2})$ 

\bea
&& 
\left\langle   e^{  -i \varphi_1/2}(z_1)  e^{  i \varphi_1(\theta+1/2)}(z_3)  e^{  -i \theta \varphi_1}(z_4)   \right\rangle 
 = z_{13}^{-{\theta\over 2}-\ft{1}{4} } z_{14}^{{\theta\over 2}}  z_{34}^{-\theta^2-\ft{\theta}{2} } \nn\\
 && 
\left\langle     e^{  -i \theta \varphi_2}(z_3)  e^{  i \theta \varphi_2}(z_4) \psi^2 (z_5)     \psi^{\bar 2} (z_6)  \right\rangle 
 =    \ft{1}{2}  z_{34}^{-\theta^2} z_{35}^{-\theta}  z_{36}^{\theta} z_{45}^{\theta}  z_{46}^{-\theta}   z_{56}^{-1}  \nn\\
 && 
\left\langle     e^{  -i  \varphi_3/2}(z_2)  e^{  i  \varphi_3/2}(z_4)  \right\rangle 
 =   z_{24}^{-\ft{1}{4}}
\eea 
 \begin{equation}
I_2=\int_{0}^{1} dx    (-x)^{-\ft{1}{2}-\ft{\theta}{2}} (x-1)^{-\theta} \mathcal{I}^{-\ft{1}{2}}[1-x]\int_{\mathbb{C}^+} \frac{d^2z \,  (x- z)^{-\theta-1/2}(x- \bar{z})^{\theta-1/2}}{  |z| (1- z)^{-\theta+1/2}(1- \bar{z})^{\theta+1/2}}
\end{equation}

\item $(M,N)=(3,\bar{3})$

\bea
&& 
\left\langle   e^{  -i \varphi_1/2}(z_1)  e^{  i \varphi_1(\theta+1/2)}(z_3)  e^{  -i \theta \varphi_1}(z_4)  \right\rangle 
 =  z_{13}^{-{\theta\over 2}-\ft{1}{4} } z_{14}^{{\theta\over 2}}  z_{34}^{-\theta^2-\ft{\theta}{2} }  \nn\\
 && 
\left\langle     e^{  -i \theta \varphi_2}(z_3)  e^{  i \theta \varphi_2}(z_4)  \right\rangle 
 =   z_{34}^{-\theta^2}  \nn\\
 && 
\left\langle     e^{  -i  \varphi_3/2}(z_2)  e^{  i  \varphi_3/2}(z_4) \psi^3 (z_5)     \psi^{\bar 3} (z_6)   \right\rangle 
 =  \ft{1}{2} z_{24}^{-\ft{1}{4}}  z_{25}^{-\ft{1}{2}}  z_{26}^{\ft{1}{2}}  z_{45}^{\ft{1}{2}}  z_{46}^{-\ft{1}{2}}  z_{56}^{-1}
 \label{4bound33}
\eea 
  \begin{equation}
I_3=\int_{0}^{1} dx    (-x)^{-\ft{1}{2}-\ft{\theta}{2}} (x-1)^{-\theta} \mathcal{I}^{-\ft{1}{2}}[1-x]\int_{\mathbb{C}^+} \frac{d^2z }{  z (1- \bar{z}) |x- z|}
\end{equation}
  
  In all these cases for a generic angle and generic complex condensate both the graviton and the B-field are emitted. 
  
\section{Comments on the Supergravity Solution}

It's very common in supergravity to find extremal solutions parametrized by arbitrary harmonic functions. The prototypical example is the Reissner-Nordstrom: in the limit in which the horizons coincide, the solution is extremal and becomes part of the larger Majumdar-Papapetrou family of solutions.
To our knowledge, even  though there is a conspicuous literature  on SUGRA  bound states of branes at angles (see for instance \cite{Breckenridge:1997ar}-\nocite{Behrndt:1997ph,Costa:1997dt,Balasubramanian:1997az}\cite{Bertolini:2000ei}) there is no extremal BPS solution expressed in terms of arbitrary harmonics functions for branes intersecting at angles. The arbitrary harmonics are crucial in order to interpret  the solutions as excitations of the  `naive' intersection of branes and to match with the stringy worldsheet computations.

To be concrete, let's remember what happen in the case of  perpendicular $D3$ branes. The `naive' solution for marginally (zero binding energy) bound states of intersecting susy branes can be obtained by the harmonic superposition rule \cite{Tseytlin:1996bh,Gauntlett:1996pb}. For instance, limiting for simplicity to the metric, the system of 4 perpendicular $D3$ brane pictured in Fig. \eqref{fig4} for $\theta=0$ is given by:

\bea
ds^2 &=& (H_0 H_1 H_2 H_3)^{-\frac{1}{2}}dt^2+(H_0 H_1 H_2 H_3)^{\frac{1}{2}}\sum_{i=1}^3dx_i^2 +(\frac{H_2 H_3}{H_0 H_1})^{\frac{1}{2}}dy_1^2+ (\frac{H_0 H_1}{H_2 H_3})^{\frac{1}{2}}d\tilde{y}_1^2+\nn \\
&+& (\frac{H_0 H_2}{H_1 H_3})^{\frac{1}{2}}dy_2^2+ (\frac{H_1 H_3}{H_0 H_2})^{\frac{1}{2}}d\tilde{y}_2^2+(\frac{H_0 H_3}{H_1 H_2})^{\frac{1}{2}}dy_3^2+ (\frac{H_1 H_2}{H_0 H_3})^{\frac{1}{2}}d\tilde{y}_3^2
\eea

where $H_A=1+\frac{Q_A}{r}$.

The associated metric parametrized in terms of  arbitrary harmonic functions has been given has been written in \cite{Lunin:2015hma,Bianchi:2016bgx}. In particular in \cite{Bianchi:2016bgx} it was given in terms of 8 harmonic functions $H_a=\{ V, L_I, K_I, M \} $:

\bea
ds^2  &=& -   e^{2U}( dt+w)^2 +e^{-2U} \,  \sum_{i=1}^3 dx_i^2 +   \sum_{I=1}^3   \left[  { d y_I^2 \over  V e^{2U} Z_I }  + V e^{2U} Z_I  \,  \tilde e_I^2  \right]
\label{bmp}
\eea

with  
  \bea
   Z_I &=& L_I +{ |\epsilon_{IJK}|\over 2} {K_J K_K\over V} \nn\\
 \mu &=& { M\over 2} +{L_I K_I \over 2\, V}+{ |\epsilon_{IJK}| \over 6} \, {K_I K_J K_K \over V^2}  \nn \\
  e^{-4U} &=&  Z_1 Z_2 Z_3 V-\mu^2  V^2 \qquad ~~~~~~~~~~ \nn \\
   b_I   &=&   \frac{K_I}{V }-\frac{\mu}{Z_I}    \qquad \qquad ~~~~~~~~~~~~~~~~~ \nn \\
   \tilde e_I &=&    d \tilde y_I   -  b_I\, dy_I \nn \\
   {*^{flat}_3}dw  &=&  V d \mu-\mu dV-V Z_I d\left( \frac{K_I}{V }\right)    
   \eea

This solution, that besides the metric includes the $C_4$ Ramond-Ramond field, reduces to the naive one if one chooses $L_I=H_I$,  $V=H_0$ and $K_I=M=0$, but in principle it can account for the hairs of the 4d black holes for a proper choice of the harmonic functions, as suggested by the matching with stringy microstates obtained in \cite{Bianchi:2016bgx}. This is precisely what has been done in the case of the D1-D5 fuzzball \cite{Lunin:2001fv}, in which the harmonic functions are such that: a)  far away from the fuzzball the metric resembles a black hole with same asymptotic charges, b) the metric has a `throat' region in which it approximates the `naive' solution, c) the fuzzball deviates strongly from the black hole geometry at the location of the would-be horizon and d) the geometry is non singular and horizonless.

In the case of  branes at angles the `naive' solution can be written, for instance specializing to the case of two $D3$ branes tilted in the first two internal torii (the $D3_0D3_3$ pair in Fig. \eqref{fig4}), as:

\bea
ds^2 &=& e^{2U_\theta} \bigg( -dt^2+\sum_{I=1}^2 dy_I^2+\sum_{I=1}^3 d\tilde{y}_I^2+ \\ 
&+& \sum_{A=1,2}\frac{Q_A}{r} \left( [(R_A)^1_{\,\, a} dy_a]^2+[(R_A)^2_{\,\, a} dy_a]^2 \right) \bigg) +e^{-2U_\theta} \bigg(  dy_3^2+\sum_{i=1}^3 dx_i^2 \bigg) \nn
\eea

\bea
F_5 &=& dt \wedge dr \wedge d\tilde{y}_3  \wedge \partial_r \bigg[  e^{4U_\theta} \bigg( \sum_{A=1,2}\frac{Q_A}{r} [(R_A)^{\tilde{1}}_{\,\, a} dy_a \wedge (R_A)^{\tilde{2}}_{\,\, a} dy_a ] - \nn \\ 
&-& (dy_1 \wedge dy_2-d\tilde{y}_1 \wedge d\tilde{y}_2) \frac{Q_1 Q_2}{r^2} \sin^2(\theta_1-\theta_2) \bigg)  \bigg] + \\
&+& g \, e^{4U_\theta}  d\theta \wedge d\phi \wedge dy_3  \wedge    \partial_r \bigg[ \sum_{A=1,2}\frac{Q_A}{r} [(R_A)^{1}_{\,\, a} dy_a \wedge (R_A)^{2}_{\,\, a} dy_a ]  \bigg]   \nn
\eea

with $a=1,\tilde{1},2,\tilde{2}$ running only in the tilted torii (we define $d\tilde{y}_1 \equiv dy_{\tilde{1}}$),  $A=1,2$ labelling the two $D3$ branes, $g$ the determinant of the metric and 

\bea
e^{2U_\theta}=\left(1+\frac{Q_1+Q_2}{r}+\frac{Q_1 Q_2}{r^2} \sin^2(\theta_1-\theta_2)  \right)^{-1/2}
\eea

\bea
R_A = \begin{pmatrix} \cos(\theta_A) & -\sin(\theta_A) & 0 &0\\
\sin(\theta_A) & \cos(\theta_A) & 0 &0 \\
 0 &0 &\cos(\theta_A) & \sin(\theta_A)  \\
 0 &0 &-\sin(\theta_A) & \cos(\theta_A)   \end{pmatrix} 
\eea

Even though the associated solution with arbitrary harmonics is still missing, we don't find any particular obstruction to it's existence. For instance after dimensional reduction on the internal $T^6$, focusing on the metric, the tilting of the internal torii translates in the presence of massless scalar fields grouped into the $h_{2,1}$  vector multiplets. Rotating the branes supersymmetrically is a continuous deformation, parametrized by some moduli fields, of a BPS system.  In particular these moduli pertain to the deformations of the complex structure and although the attractor mechanism \cite{Ferrara:1995ih,Ferrara:1996dd,Ferrara:1997tw} fixes many of them in terms of the charges of the solution, in general some fields remains undetermined and they can account for the freedom to tilt the perpendicular brane configuration while still keeping the system supersymmetric.

A possibility is that the tilted solution can be obtained by a chain of boost and dualities from the perpendicular solution \eqref{bmp}, in line with the results of \cite{Behrndt:1997ph}.


\end{itemize}

\section*{Acknowledgements}

A special thank to M. Bianchi and to J. F. Morales for suggestions and for reading the manuscript.
We would also like to thank V. Balasubramanian, M. Costa, M. Cvetic, G. Dall'Agata, S. Giusto, R. Myers, N. Zinnato  for suggestions and references.

 \begin{appendix}

  \section{Notations and Conventions}   

We will always take $\alpha'=2$. 

The index structure is the following:
\begin{itemize}
\item  $M, N, R \dots$ for 10-dimensional vector indices.
\item  $\hat{M}, \hat{N}, \hat{R} \dots$ for 6-dimensional vector indices.
\item  $ \alpha, \beta \dots$ for left spinorial indices in directions $\lbrace  t, x_1, x_2, x_3 \rbrace$
\item  $ \dot{\alpha}, \dot{\beta} \dots$ for right spinorial indices in directions $\lbrace  t, x_1, x_2, x_3 \rbrace$
\item  $ A,B,C \dots$ for $SO(1,5)$ spinor indices. Upper position means right spinor, lower position left spinor.
\item  $\mu, \nu, \rho  \dots $ for vectorial indices in extended space-time directions $\lbrace  t, x_1, x_2, x_3 \rbrace$
\item  $i, j, k \dots $ for vectorial indices in extended spatial directions $\lbrace  x_1, x_2, x_3 \rbrace$
\item  $1,\bar{1},2,\bar{2} \dots $ for complex vectorial indices. In particular $\lbrace  1,\bar{1} \rbrace$ parametrize the first internal torus $\lbrace  y_1,\tilde{y}_1  \rbrace$, whilst $\lbrace  4,\bar{4} \rbrace \rightarrow \lbrace  t,x_3 \rbrace$, $\lbrace  5,\bar{5} \rbrace \rightarrow \lbrace  x_1,x_2 \rbrace$.
\end{itemize}

The left and right spin fields of directions $\lbrace  t, x_1, x_2, x_3 \rbrace$ are respectively bosonized by the scalar field $\varphi_I$ according to the formula:

\begin{equation}
\emph{left}: \qquad S_\alpha=e^{\frac{i}{2}(\pm \varphi_4 \pm \varphi_5)} \;\;\;  \#(-)={\rm even}
\end{equation}

\begin{equation}
\emph{right}: \qquad C_{\dot{\alpha}}=e^{\frac{i}{2}(\pm \varphi_4 \pm \varphi_5)} \;\;\;  \#(-)={\rm odd}
\end{equation}

The complex fermions $ \Psi$ are built from the real ones $\psi$ via\footnote{It's convenient to treat the torus $\lbrace  4,\bar{4} \rbrace \rightarrow \lbrace  t,x_3 \rbrace$ in a different way by defining  $\Psi^4=\frac{1}{\sqrt{2}}(\psi^{0}+\psi^{3}),  \;\;  \Psi^{\bar 4}=\frac{1}{\sqrt{2}}(\psi^{0}-\psi^{3})$ in order to use $\bar{\sigma}^{\mu}={\lbrace \bar{\sigma}^0,\bar{\sigma}^1,\bar{\sigma}^2,\bar{\sigma}^3} \rbrace$. Otherwise one can use \eqref{complefermio} and add an $i$ factor in front of $\bar{\sigma}^3$ in the definition of $\bar{\sigma}^{\mu}$.}

\begin{equation}
  \Psi^I=\frac{1}{\sqrt{2}}(\psi^{2I-1}+i\psi^{2I})  \;\;\;\;  \Psi^{\bar I}=\frac{1}{\sqrt{2}}(\psi^{2I-1}-i\psi^{2I})
  \label{complefermio}
\end{equation}

and are bosonized as:

\begin{equation}
  \Psi^I=e^{i \varphi_I}
\end{equation}

In the text we will always use the notation $\psi$, even when referring to the complex fermion, as it should be clear from the context which of the two is used. Cocycle factors, needed to implement anticommutation, will be also omitted.

We follow \cite{Wess:320631} for the spinorial indices conventions. In particular we need the following formula:
 
\begin{equation} \eta_{\mu \nu}=(-,+,+,+)
\end{equation}

\begin{equation} \epsilon^{12}=\epsilon_{21}=+1 \;\;\; \epsilon^{21}=\epsilon_{12}=-1  
\end{equation}

\begin{equation} \epsilon_{0123}=-1
\end{equation}

\begin{equation}
\sigma^0= \begin{pmatrix} -1 & 0\\
0 & -1 \end{pmatrix}  \;\;\;\;\;  \sigma^1= \begin{pmatrix} 0 & 1\\
1 & 0 \end{pmatrix} 
\end{equation}

\begin{equation}
\sigma^2= \begin{pmatrix} 0 & -i\\
i & 0 \end{pmatrix}  \;\;\;\;\;  \sigma^3= \begin{pmatrix} 1 & 0\\
0 & -1 \end{pmatrix} 
\end{equation}

\begin{equation}
\sigma^0=\bar{\sigma}^0, \; \sigma^1=-\bar{\sigma}^1, \; \sigma^2=-\bar{\sigma}^2, \; \sigma^3=-\bar{\sigma}^3
\end{equation}

\begin{equation} (\sigma^{[\mu \nu]})_\alpha^{\,\,\,\,\,\beta}\equiv \frac{1}{4} ((\sigma^\mu)_{\alpha \dot{\gamma}}(\bar{\sigma}^\nu)^{\dot{\gamma}\beta}-(\sigma^\nu)_{\alpha \dot{\gamma}}(\bar{\sigma}^\mu)^{\dot{\gamma}\beta})
\label{sigma2}
\end{equation}

\begin{equation}
(\bar{\sigma}^{[\mu \nu]})^{\dot{\alpha}}_{\,\,\,\,\,\dot{\beta}} \equiv \frac{1}{4} ((\bar{\sigma}^\mu)^{\dot{\alpha}\gamma}(\sigma^\nu)_{\gamma \dot{\beta}}-(\bar{\sigma}^\nu)^{\dot{\alpha}\gamma}(\sigma^\mu)_{\gamma \dot{\beta}})
\label{sigma2b}
\end{equation}

The four dimensional gamma matrices in the Weyl basis are:

\begin{equation}
\Gamma^\mu=\begin{pmatrix} 0 & \sigma^\mu\\
\bar{\sigma}^\mu& 0 \end{pmatrix} 
\end{equation}

with associated charge conjugation matrix:

\begin{equation}
   (\Gamma^m)^T=- C (\Gamma^m) C^{-1}
\end{equation}

\begin{equation}
C=\begin{pmatrix} i (\sigma^2 \bar{\sigma}^0)^{\,\,\,\,\,b}_a & 0& \\
0 &  i (\bar{\sigma}^2 \sigma^0)^{\dot{c}}_{\,\,\,\,\,\dot{d}} \end{pmatrix}  =\begin{pmatrix} 0 & -1& 0 &0\\
1 & 0& 0 &0\\
0 & 0& 0 &1\\
0 & 0& -1 &0 \end{pmatrix} =\begin{pmatrix}
\epsilon_{ab} & 0\\ 0& -\epsilon_{\dot a \dot b}   \end{pmatrix} 
\end{equation}

    \section{OPEs and Correlators}

\begin{equation}
   \langle  \psi^{\mu}(z) \psi^{\nu}(w) \rangle= \frac{\eta^{\mu \nu}}{(z-w)} 
\end{equation}

\begin{equation}
   \langle  X^{\mu}(z) X^{\nu}(w) \rangle= - \eta^{\mu \nu} \, log(z-w) 
\end{equation}

\begin{equation}
   \langle \partial_z X^{\mu}(z) e^{ikX}(w) \rangle= -\frac{ik^{\mu}}{(z-w)} 
\end{equation}

\begin{equation}
  \langle \prod^{N}_{i=1} e^{q_i \varphi(z_i) }\rangle=  \\
 \prod^{N}_{i<j} (z_i-z_j)^{- q_i \cdot q_j}   \;\;\;\bf{ (superghost)}
\end{equation}

\begin{equation}
  \langle \prod^{N}_{i=1} e^{i\lambda_i \varphi(z_i) }\rangle=  \\
 \prod^{N}_{i<j} (z_i-z_j)^{+ \lambda_i \cdot \lambda_j}
\end{equation}

 We use the following basics OPEs in four dimensions:   
 
\begin{equation}
     C^{\dot{\alpha}} (z) S^{\beta} (w) \sim -\frac{1}{\sqrt{2 }} (\bar{\sigma}_\mu)^{ \dot{\alpha}\beta} \psi^\mu(w) 
     \label{opecs}
\end{equation}

\begin{equation}
   \psi^\mu \sim \frac{1}{\sqrt{2}} (\bar{\sigma}^\mu)^{\dot{\beta}\alpha}   C_{\dot{\beta}} S_{\alpha}
   \label{psi}
\end{equation}

\begin{equation}
  S_{\alpha}(z)  S_{\beta} (w)\sim \frac{\epsilon_{\alpha \beta}}{(z-w)^{1/2}}
  \label{opes}
\end{equation}

\begin{equation}
  C_{\dot{\alpha}}(z)  C_{\dot{\beta}}(w) \sim - \frac{\epsilon_{\dot{\alpha}\dot{\beta}}}{(z-w)^{1/2}} 
  \label{opec}
\end{equation}

\begin{equation}
   \psi^m(z) C^{\dot{\beta}} (w) \sim \frac{1}{\sqrt{2}} \frac{ (\bar{\sigma}^m)^{ \dot{\beta}\alpha} S_\alpha(w) }{(z-w)^{1/2}} 
\end{equation}

\begin{equation}
   \psi^\mu(z) S_{\alpha} (w) \sim -\frac{1}{\sqrt{2}} \frac{ (\sigma^\mu)_{\alpha \dot{\beta}} C^{\dot{\beta}}(w) }{(z-w)^{1/2}} 
\end{equation}

\begin{equation}
   \psi^\mu   \psi^\nu (z) S_{\alpha} (w) \sim  \frac{  (\sigma^{\mu\nu})_{\alpha}^{\,\,\,\,\,\beta} S_{\beta}(w) }{(z-w)} 
\end{equation}

\begin{equation}
   \psi^\mu   \psi^\nu (z) C^{\dot{\alpha}} (w) \sim  \frac{ (\bar{\sigma}^{\mu\nu})^{\dot{\alpha}}_{\,\,\,\,\,\dot{\beta}} C^{\dot{\beta}}(w) }{(z-w)} 
\end{equation}

 \subsection{How to Fix the Normalization}
 
  As an example, let's explictly fix the normalization of \eqref{opecs}.

\begin{equation}
 S^{1} =e^{\frac{i}{2} (\varphi_1+\varphi_2)}\;\;\;\;\;
S^{2}=e^{\frac{i}{2} (-\varphi_1-\varphi_2)}
\end{equation}

\begin{equation}
C^{\dot 1}= e^{\frac{i}{2} (-\varphi_1+\varphi_2)} \;\;\;\;\;
 C^{\dot 2}=e^{\frac{i}{2} (+\varphi_1-\varphi_2)}
\end{equation}

Therefore, choosing $\dot \alpha =\dot 1, \; \beta=2$ and using \eqref{complefermio}:

\begin{equation}
     C^{\dot 1} (z) S^{2} (w) \sim e^{-\varphi_1} = \Psi^{\bar 1}= \frac{1}{\sqrt{2}} (\psi^1-i \psi^2) =N (\bar{\sigma}_\mu)^{ \dot{1}2} \psi^{\mu} 
\end{equation}

So we have:

\begin{equation}
   N (\bar{\sigma}_\mu)^{ \dot{1}2} \psi^\mu = N(\psi^1  (\bar{\sigma}_1)^{ \dot{1}2}  +\psi^2  (\bar{\sigma}_2)^{ \dot{1}2} )=N(\psi^1  (-1) +\psi^2  (i) )=-N(\psi^1 -i \psi^2 )
\end{equation}

From the comparison, it follows that $N=-\frac{1}{\sqrt{2}}$.

 \section{Rotation Matrix for Tilted Branes}

Consider two parallel branes in the $(x,y)$ plane. At the beginning the first and second brane are fixed in $y=0$, that is they are Neumann in $x$ and Dirichlet in $y$, where $X,Y$ are the open string coordinates in target space. This is equivalent to $\partial_{z} X=\partial_{\bar{z}}X$ and  $\partial_{z} Y=-\partial_{\bar{z}}Y$  in complex coordinates. 
If we tilt a brane by an angle $\theta$ with respect to $y=0$, we need to impose $\partial_{z} X'=\partial_{\bar{z}}X'$ and  $\partial_{z} Y'=-\partial_{\bar{z}}Y'$ where:

\begin{equation}
\begin{pmatrix} X' \\
Y' \end{pmatrix} = \begin{pmatrix} \cos(\theta) & -\sin(\theta)\\
\sin(\theta) & \cos(\theta) \end{pmatrix}  \begin{pmatrix} X \\
Y \end{pmatrix} 
\end{equation}

After some algebra, for a counterclockwise rotation we find:

\begin{equation}
\partial_{z} X'^{\mu}=R^{\mu}_{\nu} \partial_{\bar{z}}X^{\nu}  \;\;\;\;\;\; X^{1}=X, X^2=Y
\end{equation}

\begin{equation}
R(\theta) = \begin{pmatrix} \cos(2\theta) & -\sin(2\theta)\\
-\sin(2\theta) & -\cos(2\theta) \end{pmatrix} 
\end{equation}

For a clockwise rotation we have:

\begin{equation}
R(\theta) = \begin{pmatrix} \cos(2\theta) & \sin(2\theta)\\
\sin(2\theta) & -\cos(2\theta) \end{pmatrix} 
\end{equation}

 \section{Two Boundary Amplitude}

We split the calculation into two sections: the first one dealing with the amplitude with all the fermions in extended space-time directions and the second focusing on compact directions. Here we start by computing the correlators common to both sections:

\begin{equation}
\langle \sigma^{(1)\dagger}_{1/2 -\theta}(z_1) \sigma^{(1)}_{1/2 -\theta}(z_2) \rangle \langle \sigma^{(2)}_{1/2 -\theta}(z_1) \sigma^{(2)\dagger}_{1/2 -\theta}(z_2) \rangle=z_{12}^{-\ft{1}{2} +2 \theta^2 }
\label{spin1}
\end{equation}

\begin{equation}
 \langle e^{-\ft{1}{2} \varphi}(z_1)e^{-\ft{1}{2}\varphi}(z_2) e^{-\varphi}(z_4)  \rangle=z_{12}^{-\ft{1}{4}}z_{14}^{-\ft{1}{2}}z_{24}^{-\ft{1}{2}}
 \label{super1}
\end{equation}

The $\partial X $ piece in \eqref{closedw} gives vanishing contribution due to $k^{M}G_{MN}=0$.

\subsection{Fermions in Extended Space-time  Dimensions}

The remaining internal part gives:

\begin{equation}
 \langle  e^{  i \theta\varphi_1} e^{ - i \theta\varphi_2} e^{ - i \varphi_3/2} (z_1)
 e^{  -i \theta\varphi_1} e^{  i \theta\varphi_2} e^{  i \varphi_3/2} (z_2)\rangle= z_{12}^{-2 \theta^2 -1/4}
\end{equation}

Taking  $M=\mu$ and $N=\nu$, the last correlator, apart from overall constants, is:

\begin{equation}
(k)_{\rho} (ER)_{\mu \nu}  (\bar{\sigma}^\nu)^{\dot{\gamma}\delta}   \langle S^{\alpha}  (z_1) C^{\dot{\beta}} (z_2) J^{[\rho \mu]} (z_3) C_{\dot{\gamma}} S_{\delta} (z_4) 
  \rangle
\end{equation}

in which we have defined the current $ J^{[\rho \mu ]} =:\psi^{\rho} \psi^{\mu}:$ and used the SO(1,3) relation $
   \psi^\nu \sim \frac{1}{\sqrt{2}} (\bar{\sigma}^\nu)^{\dot{\gamma}\delta}   C_{\dot{\gamma}} S_{\delta} $.
     Now we split the current in the dual (left) and anti-self dual part (right)\footnote{According to the convention $\star_4 A^{\mu \nu} \equiv \ft{\ii}{2}\epsilon^{\mu \nu \rho \sigma} A_{ \, \rho \sigma}$.}: 

\bea
 J^{[\rho \mu]} =J^{[\rho \mu]}_L+J^{[\rho \mu]}_R \nn \\
 J^{[\rho \mu]}_L=\frac{1}{2}(J^{[\rho \mu]}+\frac{i}{2} \epsilon^{\rho \mu \tau \zeta} J_{[\tau \zeta]}) \nn \\
J^{[\rho \mu]}_R=\frac{1}{2}(J^{[\rho \mu]}-\frac{i}{2} \epsilon^{\rho \mu \tau \zeta} J_{[\tau \zeta]})
\eea

Using the fact that $SO(1,3)$ Weyl spinors of different chirality have vanishing 2-point correlation function, we find:

\begin{equation}
(k)_{\rho} (ER)_{\mu \nu}  (\bar{\sigma}^\nu)^{\dot{\gamma}\delta}   [ \langle S^{\alpha}  _{(z_1)}   (J^{[\rho \mu]}_L) _{(z_3)} (S_{\delta})_{(z_4)} \rangle (-\delta^{\dot{\beta}}_ {\dot{\gamma}} z_{24}^{-1/2})+ \langle 
 C^{\dot{\beta}} _{(z_2)} ((J^{[\rho \mu]}_R )_{(z_3)} (C_{\dot{\gamma}})_{(z_4)}  \rangle(\delta^{\alpha}_ {\delta} z_{14}^{-1/2}) ]
\end{equation}

having used $\epsilon^{\alpha \beta} \epsilon_{ \beta \delta}= \delta^{\alpha}_{\delta}$ and

\bea
  S_{\alpha}(z)  S_{\beta} (w)\sim \frac{\epsilon_{\alpha \beta}}{(z-w)^{1/2}} \qquad
  C_{\dot{\alpha}}(z)  C_{\dot{\beta}}(w) \sim -\frac{\epsilon_{\dot{\alpha}\dot{\beta}}}{(z-w)^{1/2}} 
\eea

Conformal symmetry fixes the 3-point correlation functions of quasi-primary fields to be of the standard form

\begin{equation}
\langle \Phi_1(z_1) \Phi_2(z_2) \Phi_3(z_3) \rangle= \frac{\mathcal{N}_{3Pt}}{z_{12}^{h_1+h_2-h_3}z_{23}^{h_2+h_3-h_1}z_{13}^{h_1+h_3-h_2}}
\end{equation}

Where $h$ is the conformal dimension of the fields and $\mathcal{N}_{3Pt}$ is a normalization $\mathcal{N}_{3Pt}$ constant. Using \eqref{sigma2}, \eqref{sigma2b} and that $h(J)=1$, $h(S)=h(C)=1/4$, we get\footnote{For instance the first correlator, in the group theory language of $so(4)\sim so(2) \oplus so(2)$  is  $\textbf{2}_L \otimes (\textbf{4} \otimes \textbf{4})_{asd} \otimes \textbf{2}_L=\textbf{2}_L \otimes (\textbf{3}_L,\textbf{1}_R) \otimes \textbf{2}_L= (\textbf{3}_L,\textbf{1}_R) \otimes   (\textbf{1}_L+\textbf{3}_L,\textbf{1}_R)$ where $(\textbf{4} \otimes \textbf{4})_{asd}$ is the self dual antisymmetric part of the product and $\textbf{4}=(\textbf{2}_L,\textbf{2}_R)$.  Therefore the correlator has a singlet coming from  $ (\textbf{3}_L,\textbf{1}_R)\otimes   (\textbf{3}_L,\textbf{1}_R)=\textbf{1}_L+\textbf{3}_L+\textbf{5}_L$. This means that the product is symmetric in the spinor indices (if properly lowered or raised) and antisymmetric in the  vectorial indices.}:
 
\begin{equation}
\langle S^{\alpha}  _{(z_1)}(J^{[\rho \mu]}_L )_{(z_3)}  (S_{\delta}) _{(z_4)} \rangle = \frac{ (\sigma^{[\rho \mu]})_\delta^{\,\,\,\,\, \alpha}}{ z_{13}z_{14}^{-1/2}z_{34}}
\end{equation}
 
\begin{equation}
  \langle 
  C^{\dot{\beta}} _{(z_2)} (J^{[\rho \mu]}_R )_{(z_3)} (C_{\dot{\gamma}})_{(z_4)} \rangle=-\frac{(\bar{\sigma}^{[\rho \mu]})^{\dot{\beta}}_{\,\,\,\,\, \dot{\gamma}}}{ z_{23}z_{24}^{-1/2}z_{34}}
\end{equation}

having used the self-duality/anti self-duality  properties of the Lorentz generators.
Now we employ the relations

\begin{equation}
  (\bar{\sigma}^\nu)^{\dot{\beta}\delta}  (\sigma^{[\rho \mu]})_\delta^{\,\,\,\,\,\alpha}  = \frac{1}{2} [-i \epsilon^{\rho \mu \nu \eta}  (\bar{\sigma}_\eta)^{\dot{\beta}\alpha}+ \eta^{\mu \nu} (\bar{\sigma}^\rho)^{\dot{\beta}\alpha}- \eta^{\rho \nu} (\bar{\sigma}^\mu)^{\dot{\beta}\alpha}]
 \end{equation}

\begin{equation}
  (\bar{\sigma}^\nu)^{\dot{\gamma}\alpha} (\bar{\sigma}^{[\rho \mu]})^{\dot{\beta}}_{\,\,\,\,\, \dot{\gamma}} = \frac{1}{2} [-i \epsilon^{\rho \mu \nu \eta}  (\bar{\sigma}_\eta)^{\dot{\beta}\alpha}- \eta^{\mu \nu} (\bar{\sigma}^\rho)^{\dot{\beta}\alpha}+ \eta^{\rho \nu} (\bar{\sigma}^\mu)^{\dot{\beta}\alpha}]
 \end{equation}

to obtain the integrand. Indeed putting everything together and gauge fixing to  $z_1=-\infty$, $z_2=x$, $z_3=i$, $z_4=-i$ we get:

\begin{equation}
 \langle c(z_1)c(z_3) c(z_4)  \rangle=z_{13}z_{14}z_{34}
 \label{ghost1}
\end{equation}

\begin{equation}
\begin{split}
& \mathcal{A}^{NC}_{\mu\bar{\mu} NSNS} = \mu_\alpha \bar{\mu}_{\dot{\beta}} \int_{-\infty}^{+\infty} dx (z_{13}z_{14}z_{34})    z_{12}^{-\ft{1}{2} +2 \theta^2 }   z_{12}^{-\ft{1}{4}}z_{14}^{-\ft{1}{2}}z_{24}^{-\ft{1}{2}}   z_{12}^{-2 \theta^2 -1/4}\\
&  \left( \frac{i}{2 \sqrt{2}} \right)\big[(-i \epsilon^{\rho \mu \nu \eta}  (\bar{\sigma}_\eta)^{\dot{\beta}\alpha}( k)_\rho  (ER)_{\mu \nu} +(ER) (k \bar{\sigma})^{\dot{\beta}\alpha}-  (\bar{\sigma} ER k)^{\dot{\beta}\alpha})  z_{13}^{-1}z_{34}^{-1}z_{14}^{1/2}z_{24}^{-1/2}+ \\
& (-i \epsilon^{\rho \mu \nu \eta}  (\bar{\sigma}_\eta)^{\dot{\beta}\alpha}( k)_\rho  (ER)_{\mu \nu} -(ER) (k \bar{\sigma})^{\dot{\beta}\alpha}+  (\bar{\sigma} ER k)^{\dot{\beta}\alpha})z_{23}^{-1}z_{24}^{1/2}z_{34}^{-1} z_{14}^{-1/2})\big]
\end{split}
\end{equation}

Where $(ER)= (ER)_{\mu \nu} \eta^{\mu \nu}$ and $NC$ stands for non compact. The term $\bar{\sigma} ER k$ can be discarded if we are interested in a purely spatial (non internal) closed string momentum, using the fact that the momentum is transverse and the particular form of the reflection matrix in that directions, that is $kR=-k$. After some manipulations we obtain:

\begin{equation}
\mathcal{A}^{NC}_{\mu\bar{\mu} NSNS} =  \left( \frac{i}{2 \sqrt{2}} \right)  \int_{-\infty}^{+\infty}  dx      \left[ (-i \epsilon^{\rho \mu \nu \eta}  v_\eta (  k)_\rho  (ER)_{\mu \nu} )\frac{2x}{(1+x^2)} + (ER) (k v)\frac{-2i}{(1+x^2)} \right]
 \end{equation}
 
 substituting the open string condensate $
v^\rho=\mu_\alpha \bar{\mu}_{\dot{\beta}} (\bar{\sigma}^\rho)^{\dot{\beta}\alpha} $ and with $kv=k^iv^i$. The first part integrates to zero, while the second integral is simply\footnote{In all these calculations we have put the open strings momenta to zero from the beginning. If one wants to be rigorous, he should first do the full computation and then send all the open string momenta to zero. For instance the Koba-Nielsen factor, after using gauge fixing and kinematics of the momenta (and apart from an overall infinite constant due to $z_1=-\infty$) is: 

\begin{equation}
\langle e^{ik_1 X}(z_1) e^{ik_2 X} (z_2) e^{ik X}(z_3) e^{ik R X}(z_4)  \rangle=  |1+x^2|^{ \alpha' k_2   k}    |2i|^{ \alpha'  k    Rk /2}  
\end{equation}

Therefore the correct integral \eqref{pi}, using the integral beta function representation  is \cite{Stieberger:2009hq}:

\begin{equation}
\lim_{k2 \rightarrow 0} \int_{-\infty}^{+\infty}   \frac{dx}{(1+x^2)}  |1+x^2|^{ \alpha' k_2   k}   =\lim_{k2 \rightarrow 0} B(-\alpha' k_2   k+\ft{1}{2} ,\ft{1}{2} )=\pi
 \end{equation}
}: 
 
\begin{equation}
  \int_{-\infty}^{+\infty}   \frac{ dx}{(1+x^2)}= \pi
  \label{pi}
 \end{equation}

 Finally the amplitude of a massless closed string scattering on two tilted  $D3$ branes is given by:

\begin{equation}
\mathcal{A}^{NC}_{\mu\bar{\mu} NSNS} = \frac{\pi }{\sqrt{2}}   (ER)(kv) 
 \end{equation}
 
 Employing \eqref{bulkb}, one finds that the only emitted fields are the diagonal components of the metric and using 

\bea\int \frac{ d^3k}{(2 \pi)^3} \frac{-i k_j}{ |k|^2} e^{ikx}= \frac{1}{4 \pi} \frac{x_j}{ |x|^{3}}
\eea

 one finds that the scaling of the harmonic function is given by $\frac{x_j}{ r^{3}}$.

\subsection{Fermions in Compact Dimensions}

One of the fermion of the closed string vertex still needs to be in the first four directions, in order to saturate the spin fields charge. The remaining fermions must be of opposite chirality:  $(M,N)=(I,\bar{I})$  in \eqref{01closed}. We start with $(M,N)=(3,\bar{3})$ and the conjugate choice $(M,N)=(\bar{3},3)$. The correlators \eqref{spin1}, \eqref{super1} and \eqref{ghost1} are unchanged. What is left is\footnote{Notice that we have simplified the calculation by employing only scalar fields in the non compact space, otherwise we should have computed excited twist fields coming from the correlators of scalars and twist fields.
}:

\begin{equation}
(k)_\rho \langle S^{\alpha} _{(z_1)} C^{\dot{\beta}} _{(z_2)} \psi^{\rho} _{(z_3)}    \rangle =(k)_\rho \frac{1}{\sqrt{2}}\frac{(\bar{\sigma}^{\rho})^{\dot{\beta}\alpha}}{z_{13}^{1/2}z_{23}^{1/2}}
\end{equation}

\begin{equation}
 \langle  e^{  i \theta\varphi_1} e^{ - i \theta\varphi_2}  (z_1)
 e^{  -i \theta\varphi_1} e^{  i \theta\varphi_2} (z_2)\rangle= z_{12}^{-2 \theta^2}
\end{equation}
\begin{equation}
 \langle  e^{ - i \varphi_3/2}  (z_1)
 e^{  i \varphi_3/2} (z_2) e^{ \pm i \varphi_3}  (z_3)
 e^{  \mp i \varphi_3} (z_4) \rangle= \ft{1}{2} z_{12}^{-\ft{1}{4}} z_{13}^{\mp \ft{1}{2}} z_{14}^{ \pm \ft{1}{2}}  z_{23}^{\pm \ft{1}{2}} z_{24}^{\mp \ft{1}{2}} z_{34}^{-1}
\end{equation}

Summing these two contributions and gauge fixing to  $z_1=-\infty$, $z_2=x$, $z_3=i$, $z_4=-i$ we get:

\begin{equation}
\mathcal{A}^{3\bar3}_{\mu\bar{\mu} NSNS} =  \left( \frac{i}{2 \sqrt{2}} \right)  \int_{-\infty}^{+\infty}  dx       (k v) \left( \frac{(ER)_{3 \bar{3}}}{x-i} + \frac{(ER)_{ \bar{3} 3}}{x+i}\right)
 \end{equation}

that select the antisymmetric part of $(ER)_{3 \bar{3}}$, having used that $\int_{-\infty}^{+\infty}    \frac{ dx }{x-i}=\int_{-\infty}^{+\infty}    \frac{ dx \, (x+i) }{x^2+1}=i \pi$. Since the reflection matrix is diagonal with $(-1,1)$ eigenvalues in the third torus, the emitted field is the metric.

\begin{equation}
\mathcal{A}^{3\bar3}_{\mu\bar{\mu} NSNS} = \frac{-\pi }{\sqrt{2}}    (ER)_{[3 \bar{3}]} (k v) 
 \end{equation}
 
 The emitted field is therefore the off-diagonal (real) metric in the third torus and the scaling of the harmonic function is the one found previously for $NC$ components.

If instead we compute the $(M,N)=(1,\bar{1})$ we find a divergent or vanishing  integral for every angle. Indeed one finds:
 

\begin{equation}
\mathcal{A}^{1\bar1}_{\mu\bar{\mu} NSNS} \propto  \int_{-\infty}^{+\infty}  dx       ( x-i)^{-\theta-\ft{1}{2}}  ( x+i)^{\theta-\ft{1}{2}}
 \end{equation}
 
 that can be shown to be divergent using \cite{Stieberger:2015vya}: 
 
\bea
\int_{-\infty}^{+\infty}  dx       ( x-i)^{a}  ( x+i)^{b} =-\pi (2i)^{2+a+b} e^{-\pi i a} \frac{\Gamma(-1-a-b)}{\Gamma(-a)\Gamma(-b)} 
\eea

 The same reasoning applies to $(M,N)=(2,\bar{2})$.
 
 The calculation leading to \eqref{3c} has been done in \cite{Bianchi:2016bgx}.
 
\end{appendix}

%

\providecommand{\href}[2]{#2}\begingroup\raggedright\endgroup

\end{document}